\documentclass[a4paper,11pt]{article}
\usepackage{jheppub} 
\usepackage{lineno,color,slashed,changes}

\title{\boldmath Pseudo Nambu-Goldstone Boson DM with Linear Symmetry Breaking: Revisited}

\author{Jongkuk Kim} 
\author{and P. Ko}
\affiliation{Korea Institute for Advanced Study (KIAS), Seoul 02455, Republic of Korea}

\emailAdd{jkkim@kias.re.kr}
\emailAdd{pko@kias.re.kr}
\preprint{KIAS-P24002}

\abstract{
In this work, we revisit pseudo Nambu Goldstone boson (pNGB) DM model 
where global $U(1)$ dark symmetry is spontaneously broken as well as explicitly with broken by 
with linear symmetry breaking, focusing on the dark matter mass range in  
$500 {\rm GeV} \lesssim m_{\rm DM} \lesssim O(10)$ TeV. 
This model is interesting not only in its own in the context of pNGB DM, but also in the context of 
cosmological collider signatures from heavy particle mass regime without Boltzmann suppressions, 
$H \lesssim m_{\rm heavy} \lesssim 60 H$, through the chemical potential type interaction.  
After imposing perturbative unitarity, perturbativity, correct thermal relic density and constraints from colliders 
and (in)direct detection experiments,  we find that  pNGB DM mass is allowed up to $\sim 1 (10)$ TeV 
for the  dark Higgs mass $m_{H_2} = 1 (100)$ TeV for the Higgs-dark Higgs mixing $\sin \theta = 0.1$. 
We also consider the case where global  $U(1)$ dark symmetry is not spontaneously broken, where DM is
no longer pNGB.  In this case,  DM and dark Higgs masses are nearly degenerate in the range of 
a few TeV $\lesssim m_{{\rm DM}, H_2} \lesssim \sim 70$ TeV. 
Low mass region for pNGB DM and $H_2$ could be directly probed in colliders or indirect DM detections, 
whereas the heavy mass regime could be probed through the non-Gaussianity at the level of $f_{\rm NL} \sim O(0.01-10)$ if $H$ is as low as $O(1-10)$ TeV. 
}

\begin{document}
\maketitle
\flushbottom

\section{Introduction}
There are undeniable evidences of Dark Matter (DM)  in a wide range of cosmological scale.
However, the nature of DM remains a mystery since there are gravitational evidences \cite{Frenk:2012ph}.
Among various DM models, the Weakly Interacting Massive Particle (WIMP) is a well motivated DM candidate.
However, direct detection results have constrained the dark matter-nucleon scattering on most of simple popular  
WIMP-type DM models, since there is so far no clear direct detection evidence 
\footnote{If we consider a dark sector with rich structures, one can easily make WIMP type models 
that are perfectly safe from stringent bounds from DM direct detection experiments.  
See, for example, Ref.~\cite{Khan:2023uii}.}.
These constraints have compelled us to look into alternative DM models such as leptophilic DM, 
sub-GeV DM, axion DM, keV sterile neutrino DM, freeze-in DM and fuzzy DM.

In Ref.~\cite{Barger:2008jx}, the authors studied a simple extension of the SM 
scalar sector  with an additional complex scalar singlet field. 
The potential has a global $U(1)$ symmetry that is spontaneously and softly broken, it contains a viable WIMP DM 
candidate. Four different phenomenological classes were studied in Ref.~\cite{Barger:2008jx} depending on the existence of either soft (linear and/or quadratic) breaking terms  and spontaneous symmetry breaking. 
They pointed out that this model only with quadratic symmetry breaking gives rise to two-component DMs 
in the absence of spontaneous symmetry breaking. And in the existence of the linear breaking term means that the singlet  always gets the nonzero vacuum expectation 
value. 

A special case in Ref.~\cite{Barger:2008jx} is the pseudo Nambu-Goldstone boson (pNGB) DM 
with quadratic symmetry breaking only. In this framework, one can achieve  a naturally small DM-nucleon 
scattering thanks to cancellation between two CP-even Higgs bosons 
\footnote{It should be noted that this cancellation mechanism in Ref.~\cite{Gross:2017dan} is completely different 
from the cancellation mechanism for Higgs portal fermion or vector DM case \cite{Baek:2011aa,Baek:2012se,Baek:2014jga,Baek:2021hnl}.}\cite{Gross:2017dan}.
In Ref.~\cite{Gross:2017dan},  the authors considers a complex scalar $\varphi$ with  global $U(1)_X$ charge. 
This global $U(1)_X$ symmetry is assumed to be both spontaneously as well as explicitly 
broken down to a $Z_2$ symmetry through DM mass term, namely by quadratic symmetry breaking. 
Afterwards, more general explicit breaking patterns (linear, cubic and quartic breakings) have been considered, too 
\cite{Barger:2008jx, Chiang:2017nmu, Gross:2017dan, Huitu:2018gbc,Alanne:2018zjm,Azevedo:2018oxv,Karamitros:2019ewv,Cline:2019okt, Arina:2019tib,Coito:2021fgo}. 
And it was found that the cancellation mechanism rendering the tree-level direct detection scattering 
cross section be negligible works either in the case of $Z_2$ symmetric pNGB with the global $U(1)_X$ 
softly broken by a quadratic term in the potential \cite{Gross:2017dan}, or in case certain conditions are met 
among various couplings and mass parameters in the general soft breaking terms \cite{Alanne:2020jwx}. 
Taking into account one-loop corrections, DM-nucleon scattering cross section was computed in Refs.~\cite{Alanne:2018zjm, Azevedo:2018exj,Ishiwata:2018sdi}. 
It has been shown that for parameter points that satisfy the correct DM relic density, the one-loop 
cross section is below $10^{-50} {\rm cm^2}$.
DARWIN \cite{DARWIN:2016hyl}, one of the future direct detection experiments,  only gives slightly 
stronger bound compared to the perturbative unitarity condition \cite{Ishiwata:2018sdi}.
In Refs.\cite{Abe:2021nih, Cho:2022zfg}, authors studied direct detection scattering and a possibility to verify the pNGB model at International Linear Collider (ILC) considering both quadratic and linear breaking terms together.
Additionally, in general pNGB models, DM-nucleon scattering cross section at zero momentum transfer  does not 
automatically vanish \cite{Barger:2008jx, Alanne:2020jwx}. 

The nature of DM can be probed indirectly with the detection of the particle debris produced by DM annihilation
in the present Universe (indirect detections of DM).
In pNGB DM model, Fermi-LAT dwarf spheroidal (dSph) galaxies \cite{Fermi-LAT:2016uux}  provides constraints 
on DM annihilation into $b\bar{b}, W^+ W^-,ZZ$ and a pair of the SM Higgs boson.
In Refs.~\cite{Alanne:2018zjm, Arina:2019tib}, authors were evaluated the DM indirect detection scattering bounds 
by using a set of 41 and 45 dSphs from gamma-ray observations of Fermi-LAT telescope. 
The pNGB model with quadratic breaking has been studied to fit the galactic center gamma-ray and 
cosmic-ray antiproton excess \cite{Cline:2019okt}.

In this work, we revisit the pNGB DM where global $U(1)_X$ is softly and explicitly broken by linear breaking term, 
$( \varphi + \varphi^\dagger )$, where a complex scalar field $\varphi$ is charged under global $U(1)_X$ symmetry.  
This study is partly motivated by recent works on cosmological colliders~\cite{Bodas:2020yho}, where the authors 
considered the inflaton coupling to the current of heavy scalar particle ($\varphi$)  in fhe chemical potential form and global $U(1)_X$ symmetry broken (linear symmetry breaking).  In Ref.~\cite{Bodas:2020yho}, they considered 
the case without the spontaneous symmetry breaking (with $\lambda_\varphi = 0$ in Eq. (2.2) below),  
which will be discussed in Sec. 4.   
In such a case, Boltzmann suppression in the bispectrum could be avoided in certain parameter spaces of the 
underlying models.   And one of the present authors is working on the case where $U(1)_X$ is also spontaneously 
broken as well as explicitly broken by a linear term in $\varphi$ \cite{work_in_progress}, noticing that this class 
of global $U(1)_X$ models can make a good dark matter model.
For completeness, we shall consider both cases in this paper.
 
The paper is organized as follows. 
In Sec.~\ref{lin:break}, we review the general scalar potential and new interaction terms including heavy 
CP-even Higgs boson and pNGB DM.
In Sec.~\ref{dark:matter}, we study pNGB DM phenomenology including relic abundance and direct detection.
Also, we consider the stability of the potential and unitarity bounds and experimental constraints in 
Sec.~\ref{constraint}. We show numerical results in Sec.~\ref{num:result}.
We investigate the special case of $\lambda_{\varphi}=0$, corresponding to the cosmological collider 
physics with  in Sec.~\ref{Linear:lambda0} and we conclude in Sec.~\ref{cons}.

\section{Models and the Scalar potential with explicit soft symmetry breaking} \label{lin:break}
Let us consider global dark $U(1)_X$ and a complex scalar $\varphi$ with $U(1)_X$ charge equal to 
$Q_X (\varphi) = 1$.  The new scalar $\varphi$ couples to the SM particles through a Higgs portal term.
The renormalizable Lagrangian of the scalar sector (including the SM Higgs sector) with this global dark symmetry 
will be given by
\begin{eqnarray}
{\cal L_{\rm scalar}} & = & D_\mu H^\dagger D^\mu H + \partial_\mu \varphi^\dagger \partial^\mu \varphi - V_{\rm sym} ( H, \varphi)  -\Delta V_{\rm SB},
\end{eqnarray}
where
\begin{eqnarray}
V_{\rm sym} & = &  - \frac{\mu_H^2}{2} H^\dagger H + \frac{\lambda_H}{2} (  H^\dagger H  )^2 
- \frac{\mu_\varphi^2}{2} \varphi^\dagger \varphi + \frac{\lambda_\varphi}{2} (  \varphi^\dagger \varphi  )^2 
+  \lambda_{H\varphi} H^\dagger H \varphi^\dagger \varphi.
\end{eqnarray}
If there is no explicit symmetry breaking  ({\it i.e.} $\Delta V_{\rm SB} = 0$),  the Lagrangian will be invariant 
under global $U(1)_X$ transformation, 
\[
\varphi (x) \rightarrow e^{i \alpha} \varphi (x) ,  \ \ \ \alpha:{\rm real ~number~(constant)} .
\]
Then there will be a conserved vector current, 
\begin{equation}
	J^\mu_{-} =  \left[ \varphi^\dagger ( \partial^\mu \varphi ) - ( \partial^\mu \varphi^\dagger ) 
	\varphi \right] .
\end{equation}
Because $U(1)_X$ is broken spontaneously, there will appear a massless Nambu-Goldstone  boson 
($m_{\rm NGB} = 0$), which can contribute to the dark radiation in the early Universe \cite{Weinberg:2013kea}.

If the global $U(1)_X$ is explicitly broken by a nonzero $\Delta V_{\rm SB} \neq 0$, then  
the NGB acquires a nonzero mass ($m_{\rm NGB} \neq 0$), becoming pNGB. 
This pNGB can make a good cold DM (CDM) at renormalizable level \footnote{There could be global dark symmetry 
breaking dim-5 operators which would make EW scale DM decay fast. 
We shall ignore this possibility in this paper, and this issue could be resolved if we global dark symmetry is 
implmented to local dark gauge symmetry \cite{Baek:2013qwa}. }. 
There are a number of ways to break global $U(1)_X$ even at renormalizable levels: 
\begin{equation}
\label{eq:VSB}
\left[ \varphi^n + (\varphi^\dagger)^n \right] , ~~~{\rm and}~~~  i \left[ \varphi^n - (\varphi^\dagger)^n \right] , 
~~~{\rm with}~ n=1,2,3,4 ,
\end{equation}
both of which are Hermitian operators. 
In this work, we shall concentrate on a specific case, $n=1$,  
where the global symmetry is broken by the linear breaking term:  
\begin{eqnarray}
\Delta V_{\rm SB} &=&  - \frac{\mu_{\rm SB}^3}{2\sqrt{2}}
\left[ \varphi + \varphi^\dagger \right]
\end{eqnarray}
with $\mu_{\rm SB} > 0$.   Note that the second possibility $i( \varphi - \varphi^\dagger)$ can be 
brought into the above form by field redefinition, $\varphi \rightarrow i \varphi = e^{i \pi/2} \varphi$, which amounts 
to 
\[
\varphi_R \rightarrow - \varphi_I , \ \ \ \varphi_I \rightarrow \varphi_R .
\]
Note that the kinetic terms and the symmetric parts of the scalar potential remains invariant under this field 
redefinition.  Therefore we can focus on the above form, Eq. (\ref{eq:VSB}), without loss of generality. 
In general, one can make field redefinition, $\varphi \rightarrow e^{i \alpha} \varphi$ with 
\[
\alpha = \frac{\pi}{n} \left( \frac{1}{2} + 2 m \right) , \ \ m:{\rm integer} ,
\]
under which the first form changes into the second form in Eq. (\ref{eq:VSB}). 
Therefore it suffices to consider the first form of the symmetry breaking for simplicity \footnote{Strinctly speaking,
this is true only if there are symmetry breaking operators with only one type (namely, only one $n$). 
If there are two types of symmetry breaking terms simultaneously, one can not make field redefinitions to 
remove the 2nd type operators. }. 

The potential including the linear breaking is invariant under the following transformation: 
\begin{eqnarray}
	\varphi \to \varphi^\dagger ,
\end{eqnarray}
Under this transformation, DM ($\sim {\rm Im} \varphi$) will carry a negative parity,  which is  
an accidental symmetry, guaranteeing the stability of DM in this model at renormalizable level.  
Note that this type of explicit $U(1)_X$ symmetry breaking can generate some large signatures in cosmological 
collider framework \cite{Bodas:2020yho}, 
which is one of the main motivations to revisit here this particular pNGB DM models with linear symmetry breaking.

We assume that $\mu_H^2 >0$ so that the EW symmmetry is spontaneously broken by the
nonzero VEV of the SM Higgs doublet, $\langle H \rangle = ( 0, v_H )^T / \sqrt{2}$. 
On the other hand, we consider $\mu_\varphi^2$ can have either sign so that the $U(1)_X$ symmetry is either preserved for $\langle \varphi \rangle = 0$, or spontaneously broken for the nonzero $\varphi$ VEV, $\langle \varphi \rangle = v_\varphi / \sqrt{2}$.

\section{Linear breaking with nonzero quartic interaction of $\varphi$ ($\lambda_\varphi >0$)}
\subsection{Particle Mass Spectra and Interactions}

Let us expand the $\varphi (x)$ as 
\begin{eqnarray}
	\varphi (x) = \frac{1}{\sqrt{2}} \left[ v_\varphi + \varphi_R (x) + i \varphi_I (x) \right] \equiv 
	\frac{1}{\sqrt{2}} \left[ v_\varphi + \rho (x) + i \eta (x) \right] ,
\end{eqnarray}
where $\varphi_R \equiv \rho$ and  $\varphi_I \equiv \eta$ are real scalar fields 
\footnote{We assume $v_\eta = 0$ in this work.}, 
and expanding the SM Higgs doublet in the unitary gauge, 
\begin{eqnarray}
	H(x) = ( 0, (v_H + h(x))/\sqrt{2} )^T .
\end{eqnarray}

The ground state is defined by the vanishing tadpole conditions:
\begin{eqnarray}
	\mu_H^2 & = & \lambda_H v_H^2 + \lambda_{H\varphi} v_\varphi^2  , \\
	\mu_\varphi^2 & = & -\frac{\mu_{\rm SB}^3 }{v_\varphi} + \lambda_{H\varphi} v_H^2 + \lambda_\varphi v_\varphi^2  \label{tadpole:muPhi2}.
\end{eqnarray}
Note that the linear breaking term with nonzero $ \mu_{\rm SB}^3$ always makes  $v_\varphi$ nonzero 
($v_\varphi \neq 0)$, whether or not $U(1)_X$ symmetry is spontaneously broken ($\mu_\varphi^2 > 0$).
In order that spontaneous symmetry breaking is realized, it is important to have degenerate vacua when the  
explicit symmetry breaking is turned off.

The second order derivatives are given by the following $3\times 3$  matrix:
\begin{align}
	\partial_i \partial_j V &= 
 \begin{pmatrix}
	\frac{1}{2} \left( 3\lambda_H v^2_H +\lambda_{H\varphi} v^2_\varphi - \mu^2_H \right) & \lambda_{H\varphi} v_H v_\varphi & 0 \\
	\lambda_{H\varphi} v_H v_\varphi & \frac{1}{2} \left( \lambda_{H\varphi} v^2_H +3\lambda_{\varphi} v^2_\varphi -\mu^2_\varphi  \right) & 0\\
	0 & 0 & \frac{1}{2}\left(\lambda_{H\varphi} v^2_H +\lambda_{\varphi} v^2_\varphi -\mu^2_\varphi  \right)
\end{pmatrix} \nonumber\\
&=
 \begin{pmatrix}
\lambda_H v_H^2 & \lambda_{H\varphi} v_H v_\varphi & 0 		\\
\lambda_{H\varphi} v_H v_\varphi  & m_\eta^2 +  \lambda_\varphi v_\varphi^2 & 0\\
0 & 0 & m_\eta^2 
\end{pmatrix} 	\label{deldelV0}
\end{align}
in the $(h,\rho,\eta)$ basis.  
The mass of the pseudo scalar boson $\eta$ is given by   
\begin{eqnarray}
	m_\eta^2 = \frac{\mu_{\rm SB}^3}{2 v_\varphi}. \label{etaMass}
\end{eqnarray}
This matrix corresponds to the squared mass matrix $\mathcal{M}^2$ in the interaction basis.
Here the squared mass matrix is non-diagonal between the two scalar fields $h,\rho$. 
Considering non-zero VEV of $v_\varphi$, $\lambda_{H\varphi}$ leads to a mixing between two CP-even interaction eigenstates $(h, \varphi)$.
Thus, new scalar field $\varphi$ can communicate with the SM particles  through the Higgs portal coupling 
$\lambda_{H\varphi}$. Note that there is $m_\eta^2$ term in the $(2,2)$ component in Eq. (\ref{deldelV0}), which is 
characteristically distinct from the quadratic symmetry breaking case \cite{Gross:2017dan}.

The mixing angle $\theta$ between two CP-even Higgs bosons satisfies the relation below:
\begin{eqnarray}
	\tan 2\theta  & = & \frac{2\lambda_{H\varphi} v_H v_\varphi }{ \lambda_\varphi v^2_\varphi + m^2_\eta -\lambda_H v^2_H }.
\end{eqnarray}
One can define the mass eigenstates, $H_{1,2}$, which are
linear combinations of $h$ and $\rho$:
\begin{eqnarray}
	m_{H_{1,2}} & = & \frac{1}{2} \left[ \lambda_\varphi v^2_\varphi+ m^2_\eta +\lambda_H v^2_H \mp \sqrt{\left( \lambda_\varphi v^2_\varphi + m^2_\eta -\lambda_H v^2_H \right)^2 + 4 \left( \lambda_{H\varphi} v_H v_\varphi \right)^2 }\right].
\end{eqnarray}
Given the discovery of the SM Higgs at the LHC \cite{ATLAS:2012yve,CMS:2012qbp,ParticleDataGroup:2022pth}, we identify $H_1$ as the SM Higgs boson with
\begin{eqnarray}
	m_{H_1}=125.25 {\rm GeV}, \quad v_H=246 {\rm GeV}.
\end{eqnarray}
After the global $U(1)$ symmetry breaking, the remnant global symmetry is $Z_2$ symmetry. 
In other words, $\eta \rightarrow - \eta$ is an accidental symmetry of this model\footnote{You can find more detailed 
discussions on local $Z_2$ scalar/fermion DM models \cite{Baek:2014kna, Ko:2019wxq, Baek:2020owl, 
Baek:2022ozm}. }. 
Therefore $\eta$ can serve as a good CDM candidate as long as we ignore gravity-induced higher dimensional operators, such as 
\begin{eqnarray}
	\frac{O(1)}{M_{\rm Planck}} \times \eta F_{\mu\nu}F^{\mu\nu} ,
\end{eqnarray}
whose scale is set presumably by Planck scale. 

Thanks to a nonzero Higgs portal coupling ($\lambda_{H\varphi}\neq 0$), there are new interactions between 
CP-even Higgses and the SM particles. The interaction between $H_{1,2}$ and SM fermions $f$ are given by
\begin{eqnarray}
\mathcal{L}_{\rm H_i ff} &=& - \left(H_1 \cos\theta +H_2 \sin\theta \right) \sum_f \frac{m_f}{v_H} \bar{f}f,
\end{eqnarray} 
where $m_f$ denotes the mass of the fermion $f$.

The interaction between $H_{1,2}$ and $W,Z$ gauge bosons are given by
\begin{eqnarray}
\mathcal{L}_{\rm H_i VV} &=& m_W g \left(  W^{+\mu} W^{-}_\mu + \frac{Z^\mu Z_\mu}{2\left( 1-\sin^2\theta_W\right) }  \right)\left(H_1 \cos\theta +H_2 \sin\theta \right),
\end{eqnarray} 
where $g$ is the $SU(2)_L$ gauge coupling constant and $\sin\theta_W$ is the Weinberg or weak mixing angle. 

Now let us analyze this model in detail, starting from the particle spectra of the renormalizable parts 
of the model Lagrangian. Vanishing tadpole conditions determine the VEV's of $\varphi_R$, $v_\varphi$. 
We can replace the parameter $\mu^2_H, \mu^2_\varphi, \lambda_H, \lambda_\varphi, \mu_{SB}$ and $\lambda_{H\varphi}$ appearing in the potential with physical parameters $m_{H_1},m_{H_2}, m_\eta, v_H$ and $v_\varphi$.
Note that $H_1=h\cos\theta - \varphi\sin\theta$ is mostly SM Higgs-like state because we are interested in the small $\theta$ case. 
Now we can always consider $\theta \ll \pi/4$. 
The mass matrix in the basis $(h, \varphi)$ can be expressed in terms of the physical parameters as follows:
\begin{eqnarray}
	\begin{pmatrix}
		\lambda_H v_H^2 & \lambda_{H\varphi} v_H v_\varphi \\
		\lambda_{H\varphi} v_H v_\varphi  & m_\eta^2 +  \lambda_\varphi v_\varphi^2
	\end{pmatrix}
	=
	\begin{pmatrix}
		m^2_{H_1} \cos^2\theta + m^2_{H_2}\sin^2\theta & ~~\left( m^2_{H_2} - m^2_{H_1}\right) \sin\theta\cos\theta \\
		\left( m^2_{H_2} - m^2_{H_1}\right) \sin\theta\cos\theta&~~  m^2_{H_1} \sin^2\theta + m^2_{H_2} \cos^2\theta
	\end{pmatrix}. \label{mass:matrix}
\end{eqnarray} 
Notice that $\lambda_\varphi$ can be negative value if DM mass is approximately larger than $m_{H_2}$.

Considering the model with the linear breaking, the couplings of $H_{1,2}$ to DM become 
\begin{eqnarray}
\mathcal{L} &=&  \frac{1}{2}\left( -\lambda_{H\varphi} v_H \cos\theta + \lambda_\varphi v_\varphi \sin\theta\right) \eta^2 H_1 - \frac{1}{2}\left( -\lambda_{H\varphi} v_H \cos\theta - \lambda_\varphi v_\varphi \sin\theta \right)\eta^2 H_2 \nonumber\\
&=& \lambda_{\eta\eta H_1} \frac{\eta^2}{2} H_1 + \lambda_{\eta\eta H_2} \frac{\eta^2}{2} H_2,
\end{eqnarray}
where the dimensionful couplings $\{\lambda_{\eta\eta H_1},\lambda_{\eta\eta H_2}\}$ are
\begin{eqnarray}
\lambda_{\eta\eta H_1} &=& \left(-\lambda_{H\varphi} v_H \cos\theta +\lambda_{\varphi} v_\varphi \sin\theta \right)\nonumber\\
&=& \frac{1}{v_\varphi} \left[-\left( m^2_{H_2} - m^2_{H_1}\right) \sin\theta\cos^2\theta +\sin\theta\left( m^2_{H_1} \sin^2\theta + m^2_{H_2} \cos^2\theta -m^2_\eta \right)  \right] \nonumber\\
&=& \frac{m^2_{H_1}-m^2_\eta}{v_\varphi}\sin\theta, \\
\lambda_{\eta\eta H_2} &=& \left(-\lambda_{H\varphi} v_H \sin\theta -\lambda_{\varphi} v_\varphi \cos\theta \right)\nonumber\\
&=& \frac{1}{v_\varphi} \left[-\left( m^2_{H_2} - m^2_{H_1}\right) \sin^2\theta\cos\theta -\cos\theta\left( m^2_{H_1} \sin^2\theta + m^2_{H_2} \cos^2\theta -m^2_\eta \right)  \right] \nonumber\\
&=& -\frac{m^2_{H_2}-m^2_\eta}{v_\varphi} \cos\theta.
\end{eqnarray}

If we consider the case $\mu_\varphi^2 < 0$, namely without spontaneous symmetry breaking, the expressions 
for $H_1, H_2, \eta$ masses and their couplings are exactly the same with Eqs. (\ref{deldelV0}) that were obtained for 
the case of spontaneous symmetry breaking with $\mu_\varphi^2 > 0$.  Only difference is the sign of 
$\mu_\varphi^2$ that appears in the vanishing tadpole condition, Eq. (\ref{tadpole:muPhi2}). 
In the previous literature, only specific parameter space regions were considered ignoring this possibility. 

\subsection{Dark matter phenomenology } \label{dark:matter}
In this section, we study pNGB DM phenomenology with linear breaking mechanism.
For numerical studies, we make a model file by using FeynRules \cite{Alloul:2013bka}.

\subsubsection{Relic density of DM}
Now we will concentrate on DM thermal production.
The evolution of the DM number density $n_\eta$ is governed by the Boltzmann equation,
\begin{eqnarray}
\frac{dn_\eta}{dt}+3H n_\eta &=& -\langle \sigma v\rangle \left(n^2_\eta -\left( n^{\rm eq }_\eta\right)^2 \right),
\end{eqnarray}
where $\langle \sigma v\rangle$ is the total 2-to-2 thermal-averaged cross section.
To fit the total observed DM relic density, we require that 
\begin{eqnarray}
\Omega h^2 &\simeq& \frac{ 8.77\times 10^{-11} {\rm GeV^{-2}} x_f }{\sqrt{g_*}\langle\sigma v\rangle }.
\end{eqnarray}
where $x_f=m_\eta/T_{f}$ is the dimensionless parameter related to the freeze-out temperature. 

The important DM annihilation channels relevant to DM thermal relic density are 
$\eta\eta \to W^+W^-, ZZ$, $f\bar{f}$ and $H_i H_j$ with $i,j =1,2$. 
The thermal averaged cross section $\eta\eta\to W^+W^-$ is
\begin{eqnarray}
\langle \sigma v \rangle_{\eta\eta\to W^+W^-} &=& \frac{1}{16\pi s}\sqrt{\frac{s-4m^2_W}{s-4m^2_\eta}  } \overline{\left\vert \mathcal{M} \right\vert^2} ,
\end{eqnarray}
where $s \equiv m_{\eta\eta}^2$, and 
\begin{eqnarray}
\overline{\left\vert \mathcal{M} \right\vert^2} &=& \left\vert \frac{ \lambda_{\eta\eta H_1} g m_W \cos\theta }{s- m^2_{H_1} + im_{H_1} \Gamma_{H_1}} -\frac{  \lambda_{\eta\eta H_2} g m_W \sin\theta }{s- m^2_{H_2}+im_{H_2} \Gamma_{H_2}} \right\vert^2 \left(\frac{s^2-4m^2_W s+12m^4_W}{4m^4_W}\right) .
\end{eqnarray}
The thermal averaged cross section for the DM pair annihilation  into a pair of $Z$ is given by 
\begin{eqnarray}	
\langle \sigma v \rangle_{\eta\eta\to ZZ} &=& \frac{1}{32\pi s}\sqrt{\frac{s-4m^2_Z}{s-4m^2_\eta}  } \overline{\left\vert \mathcal{M} \right\vert^2} ,
\end{eqnarray}
with
\begin{eqnarray}
\overline{\left\vert \mathcal{M} \right\vert^2} &=& \left\vert \frac{ \lambda_{\eta\eta H_1} g_Z m_W \cos\theta }{s- m^2_{H_1} + im_{H_1} \Gamma_{H_1}} -\frac{ \lambda_{\eta\eta H_2} g_Z m_W \sin\theta  }{s- m^2_{H_2}+im_{H_2} \Gamma_{H_2}} \right\vert^2 \left(\frac{s^2-4m^2_Z s+12m^4_Z}{4m^4_Z}\right)
\end{eqnarray}
The thermal averaged cross section $\eta\eta\to f\bar{f}$ is
\begin{eqnarray}
\langle \sigma v \rangle_{\eta\eta\to f\bar{f}} &=&   \frac{1}{8\pi s}\sqrt{\frac{s-4m^2_f}{s-4m^2_\eta}  } \frac{m^2_f(s-4m^2_f)}{v^2_H} \left\vert \frac{ \lambda_{\eta\eta H_1} \cos\theta }{s- m^2_{H_1} + im_{H_1} \Gamma_{H_1}} -\frac{ \lambda_{\eta\eta H_2}\sin\theta }{s- m^2_{H_2}+im_{H_2} \Gamma_{H_2}} \right\vert^2.\nonumber\\
\end{eqnarray}
With two CP-even Higgs mediators, DM annihilation cross section is resonantly enhanced when $m_\eta \sim m_{H_{1,2}}/2$.
Near the CP-even Higgs resonances, DM annihilation cross section should be suppressed to obtain 
the correct relic density.
Away from these resonances, large values of $v_H/v_\varphi$ generally overproduce the DM relic density.
In addition, DM annihilation into $H_1 H_1,H_1H_2, H_2H_2$ final states is also possible through combined with 
Higgs boson exchanges in the $s-$channel, DM exchange in the $t-$and $u-$channels  and contact interactions. 
Taking into account all of the DM annihilation channels and using the micrOMEGAs code \cite{Belanger:2018ccd}, 
we can obtain the correct relic density.

\subsubsection{DM direct detection}
In the linear breaking scenario, DM-nucleon scattering does not have cancellation due to additional $m_\eta$ terms.
The tree-level direct detection scattering amplitude is
\begin{eqnarray}
	\mathcal{M} &=& \sin\theta\cos\theta \frac{f_N m_N}{v_H v_\varphi} \left( \frac{ m^2_{H_1} - m^2_\eta }{t- m^2_{H_1}} -\frac{m^2_{H_2} - m^2_\eta}{t- m^2_{H_2}}  \right) \bar{u}_N u_N \\
	&\simeq& \sin\theta\cos\theta \frac{f_N m_N m^2_\eta }{v_H v_\varphi} \frac{m^2_{H_2}-m^2_{H_1}}{m^2_{H_1}m^2_{H_2}} \bar{u}_N u_N,
\end{eqnarray}
where $t \equiv q^2$ is the squared 4-momentum transfer to the nucleon, $m_N$ is the nucleon mass, 
and  $f_N=0.327$ is the nucleon $\sigma$ term \cite{Young:2009zb, Crivellin:2013ipa}.
Note that this amplitude is not proportional to $t$ as in the pNGB DM models with quadratic symmetry breaking \cite{Gross:2017dan}.  It only vanishes when the two scalars $H_{1,2}$  are degenerate \cite{Abe:2021nih}. 
In Ref.~\cite{Alanne:2020jwx}, the authors mentioned that the spin-independent scattering 
at tree-level is suppressed for $U(1)_X$-invariant interactions or if there only exists quadratic term. 
In our case, it does vanish only when $m_{H_1}$ and $m_{H_2}$ are degenerate, as in Higgs portal fermion or vector DM cases \cite{Baek:2011aa,Baek:2012se,Baek:2014jga,Baek:2021hnl}. 
 
The spin-independet scattering mediated by two CP-even Higgs is 
\begin{eqnarray}
\sigma_{\rm SI} &=& \frac{f^2_N}{4\pi } \sin^2\theta\cos^2\theta \frac{m^4_N m^4_\eta}{m^4_{H_1}m^4_{H_2} v^2_H v^2_\varphi}\frac{\left( m^2_{H_2} -m^2_{H_1} \right)^2}{\left(m_N+m_\eta\right)^2}.
\end{eqnarray}
Note that in the heavy $m_{H_2}$ limit, scattering is independent of $m_{H_2}$.
Currently the most stringent DM direct detection bound comes from XENONnT experiment \cite{XENONCollaboration:2023orw}.
We adopt the upper bound and constrain our parameter space. 

\subsection{Constraints}\label{constraint}
In this Subsection, we describe the set of constraints relevant to our study.

\subsubsection{Stability of the scalar potential and Unitarity}
In order to guarantee the vacuum stability, the potential should follow conditions below 
\cite{Azevedo:2018oxv, Alanne:2020jwx}.
\begin{eqnarray}
	\lambda_H >0,\quad\quad   \lambda_{\varphi} >0, \quad\quad  \lambda_{H\varphi} > -\sqrt{\lambda_H \lambda_\varphi}, \nonumber\\
	\lambda_\varphi \sqrt{\lambda_H} +\lambda_{H\varphi} \sqrt{\lambda_\varphi} + \sqrt{\lambda_\varphi \left(\lambda_{H\varphi} +\sqrt{\lambda_H \lambda_\varphi} \right)^2} >0, \nonumber\\
	\vert\lambda_H\vert \leq 8\pi, \quad\quad \vert \lambda_{H\varphi} \vert \leq 8\pi, \quad\quad \vert\lambda_\varphi\vert \leq 8\pi.
\end{eqnarray}
From Ref. \cite{Barger:2008jx}, the stability condition of the scalar potential is also given by
\begin{eqnarray}
	\lambda_H \left( \lambda_\varphi + \frac{m^2_\eta}{v^2_\varphi}\right) > \lambda^2_{H\varphi}.
\end{eqnarray}
Note that $\lambda_\varphi$ can be negative when DM mass becomes larger in Eq. (\ref{mass:matrix}), 
in which case the vacuum stability would be ruined. 
Therefore, in this model with linear breaking, heavy DM mass is disfavored by the stability of the scalar potential 
\footnote{That $\lambda_\varphi$  can be negative when DM mass is greater than $m_{H_2}$ in Eq. (3.31), and  
the vacuum stability could be ruined was overlooked in earlier studies to our best knowledge.}. 

Also we consider the perturbative unitarity bounds from the $H_2 H_2 \to H_2 H_2$ elastic scattering 
process at high energies \cite{Chen:2014ask}: 
\[
\lambda_\varphi \lesssim8\pi/3 .
\]

\subsubsection{Higgs invisible decay}
Through the Higgs portal $\lambda_{H\varphi}$ interaction, two CP-even Higgs bosons can mix with each other.
In the CP-even mass eigenstates, $H_1$ and $H_2$, we identify $H_1$ as the 125 GeV SM Higgs-like scalar 
boson observed at CERN, and its coupling are scaled by $\cos\theta$ relative to those of the SM Higgs boson.
The lastest measurement of the Higgs decay width is $\Gamma_{H_1}=3.2^{+2.4}_{-1.7}$ MeV  \cite{ParticleDataGroup:2022pth}.   In this paper, we assume that $H_2$ is heavier than the SM Higgs bosons, 
motivated by the cosmological collider physics \cite{Bodas:2020yho}.  
Then, the SM Higgs boson is kinematically allowed to decay into a pair of DM particles when 
$m_{H_1} > m_\eta/2$.  The Higgs invisible decay bound is expressed by
\begin{eqnarray}
	{\rm Br}\left(H_1 \to \eta\eta \right) &=& \frac{\Gamma\left(H_1 \to \eta\eta \right)}{\Gamma^{\rm SM}_{H_1} + \Gamma\left(H_1 \to \eta\eta \right)}
\end{eqnarray}
with 
\begin{eqnarray}
	\Gamma\left(H_1 \to \eta\eta \right) &=& \frac{ \lambda^2_{\eta\eta H_1 }}{32\pi m_{H_1}}\left( 1- \frac{4m^2_\eta}{m^2_{H_1}}\right)^{1/2}.
\end{eqnarray}
The combined current experimental bound provided by the ATLAS and CMS \cite{ParticleDataGroup:2022pth} is
\begin{eqnarray}
	{\rm Br}(H_1\to {\rm inv.}) <  0.11.
\end{eqnarray}
We adopt the upper limit from the the ATLAS and CMS experiments.

\subsubsection{Dark Matter indirect detection}
Cosmic ray fluxes of $\gamma, e^+,$ and $\bar{p}$ generated from the cascade decays of the DM annihilation
 induced SM particle pairs are proportional to the total DM annihilation cross section $ \langle\sigma v\rangle$ 
 for a fixed  $m_{\rm DM}$ and a given set of related astrophysical parameters.
We will consider the upper limit on DM annihilation cross section based on four different observation data.

Ionizing particles coming from DM annihilations can modify the ionization history of the hydrogen and helium gases, thereby perturbing CMB anisotropies.
Planck measurement of the CMB anisotropy is utilized to get the 95\% confidence level upper limit on $ \langle\sigma v\rangle$ for given DM mass \cite{Planck:2015fie, Planck:2018vyg}.

Dwarf spheroidal galaxies are good targets to investigate DM annihilation signals thanks to the high mass-to-light ratios and low astrophysical backgrounds.
Thus, the $\gamma$-ray data in the energy range $500 {\rm MeV} - 500 {\rm GeV}$ obtained from the Fermi-LAT observation provides strong constraints on $(m_{\rm DM}, \langle\sigma v\rangle )$ parameter space \cite{Fermi-LAT:2015att}.

Positrons induced by DM annihilations receive diffusion and energy losses during its propagation through the Milky Way halo. 
These positrons can have an effect on the AMS-02 observation of the cosmic-ray positron flux in the energy range $500 {\rm MeV} - 1 {\rm TeV}$ \cite{AMS:2019rhg}. 

Lastly, H.E.S.S is sensitive to high energy gamma-rays. 
Therfore, for $m_{\rm DM} > 200 {\rm GeV}$, the constraints obtained from the H.E.S.S gamma-ray observation from the Galactic halo are stronger than those coming from other data \cite{HESS:2022ygk}.

In this work, we will testify our model taking into account DM annihilation cross section limits obtained from Planck, Fermi-LAT dSph observation, AMS-02  and H.E.S.S. and \cite{Funk:2013gxa,Lefranc:2016srp, Cuoco:2017iax, Roszkowski:2017nbc, Boddy:2018qur,Leane:2018kjk,Hoof:2018hyn,Boddy:2019qak, Montanari:2022buj}.

\subsection{Results} \label{num:result}
In this Section, we discuss pNGB DM phenomenology with linear symmetry breaking, including various conditions 
and constraints that we described in the previous subsections, and show the corresponding results.

\begin{figure}
\centering
\includegraphics[width=0.48\linewidth]{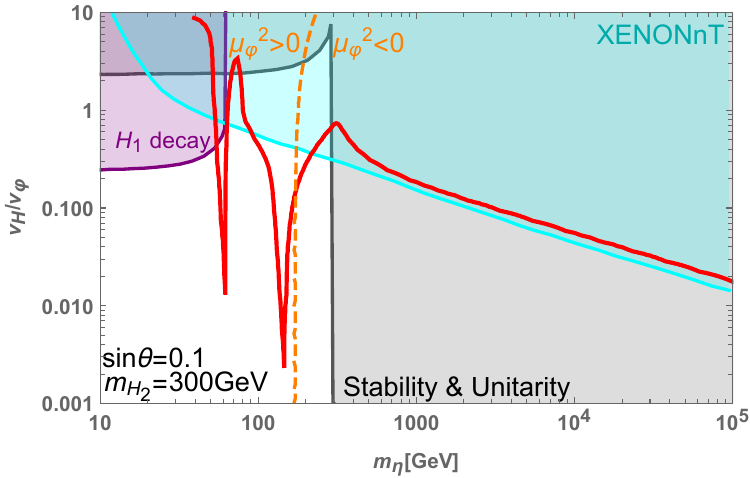}
\includegraphics[width=0.48\linewidth]{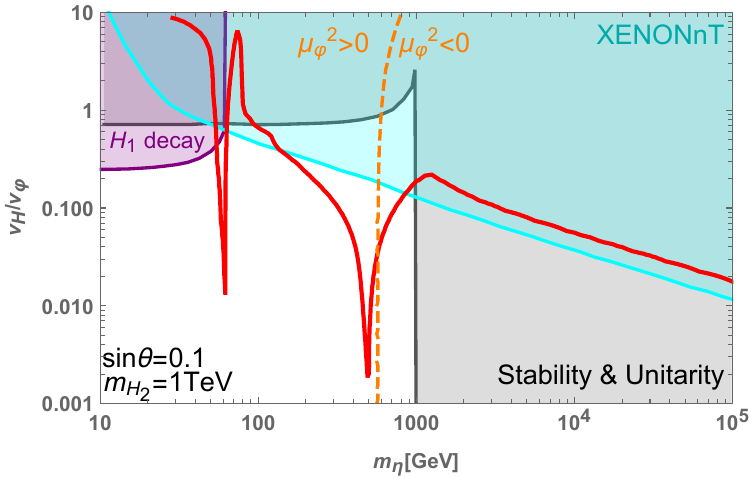}
\includegraphics[width=0.48\linewidth]{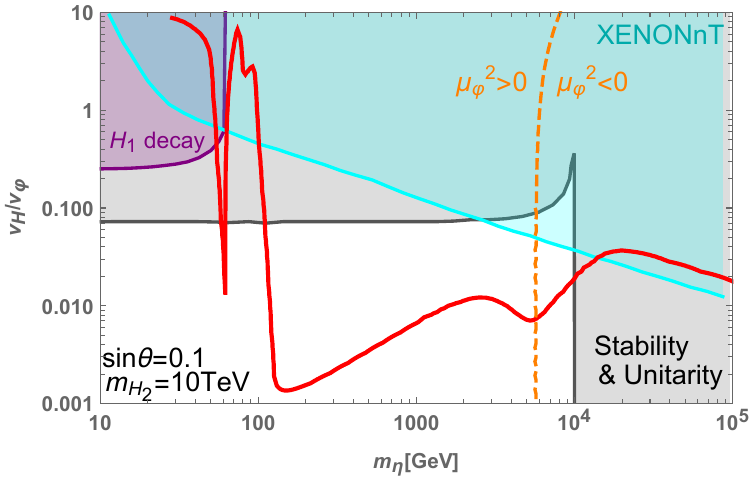}
\includegraphics[width=0.48\linewidth]{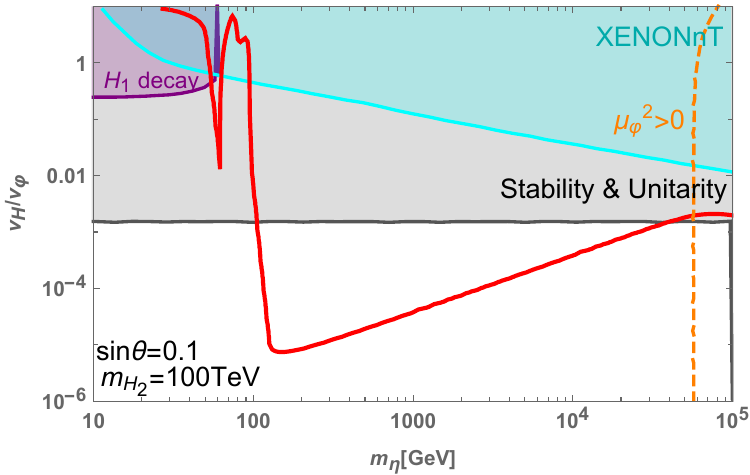}
\hspace{0.5cm}
\caption{ Allowed region for DM mass $m_\eta$ vs $v_H/v_\varphi$. We take $\sin\theta=0.1$. From top-left to bottom-right panels, we have fixed $m_{H_2}=3\times 10^2, 10^3, 10^4, 10^5$ GeV, respectively. The red line corresponds to the DM relic abundance which is consistent with the Planck data,  $\Omega h^2=0.12 \pm 0.001$ at $1\sigma$ C.L.. The purple region is ruled out by the upper bound on the Higgs invisible decay width, 
$Br(H_1 \to {\rm inv.}) < 0.11$.  
The gray area is excluded by vacuum stability and perturbative unitary conditions.  
The orange dashed line corresponds to $\mu_\varphi^2 = 0$. In the left (right) side of this orange line, one has
$\mu_\varphi^2 > 0$ $(\mu_\varphi^2 < 0)$, namely with (without) spontaneous breaking of $U(1)_X$. 
Lastly, the cyan region is ruled out by DM direct detection bound from XENONnT.  } \label{metaVSvphi}
\end{figure}
In Fig.~\ref{metaVSvphi}, we show the allowed paramter space in the $(m_\eta, v_H/v_\varphi)$ plane where we set 
the mixing angle $\sin\theta=0.1$ and $m_{H_2}=3\times 10^2,~10^3,~10^4,~10^5$ GeV, respectively.
With two CP-even neutral scalar mediators, DM annihilation cross section can be resonantly enhanced when $m_\eta \sim m_{H_{1,2}}/2$.
The first resonance effect comes from the SM Higgs boson $H_1$. 
The other is originated from the heavy CP-even Higgs boson $H_2$.
When the mass of $H_2$ is much larger than $1$TeV, resonance enhancement becomes weaker because of 
the large decay width of $H_2$.    The gray region is excluded by either the validity of the perturbative calculations 
or the stability of the scalar potential. 
When DM mass is roughly heavier than the mass of new CP-even Higgs boson $m_\eta > m_{H_2}$, 
$\lambda_{\varphi}$ becomes negative, which is not the case for the quadratic symmetry breaking. 
Therefore, there is a huge difference between our model with linear symmetry breaking 
and the pNGB model with quadratic symmetry breaking \cite{Gross:2017dan}.
The parameter regions $m_\eta > m_{H_2}$, $\lambda_{\varphi} \simeq 0$ are not allowed in 
our pNGB DM models with linear symmetry breaking because of vacuum stability conditions,  whereas they are 
allowed in the pNGB DM models with quadratic symmetry breaking .
The red solid line corresponds to the DM relic density  that is consistent with the value observed by Planck satellite \cite{Planck:2018vyg}.
The purple area is ruled out by the upper bounds on the SM Higgs invisible branching ratios \cite{ParticleDataGroup:2022pth}. 
The  pNGB DM model with quadratic symmetry breaking is not strongly constrained by 
the DM direct detection bounds because the tree-level amplitude is proportional to 
momentum transfer $t$ which is very small. 
However, in our model with linear symmetry breaking, tree level amplitude is no longer 
proportional to the momentum transfer $t$, but to $(m_{H_1}^2 - m_{H_2}^2)$, as discussed in Sec. 3.2.2.  
There appears a different kind of cancellation mechanism in the limit $m_{H_2} \rightarrow m_{H_1}$, 
just like the Higgs portal singlet fermion or vector DM models \cite{Baek:2011aa,Baek:2012se}. 
In Fig.~\ref{metaVSvphi}, cyan area is excluded by dark matter direct detection scattering bound from XENONnT 
experiments \cite{XENONCollaboration:2023orw}.  
As explained above, this bound can be relaxed only when two CP-even scalar bosons becomes degenerate.
The orange dashed line in Fig.~1 represents $\mu_\varphi^2 = 0$. The left side of the orange line corresponds to 
the region $\mu_\varphi^2 > 0$, where $U(1)_X$ is broken spontaneously.  
\begin{figure}
\centering
\includegraphics[width=0.48\linewidth]{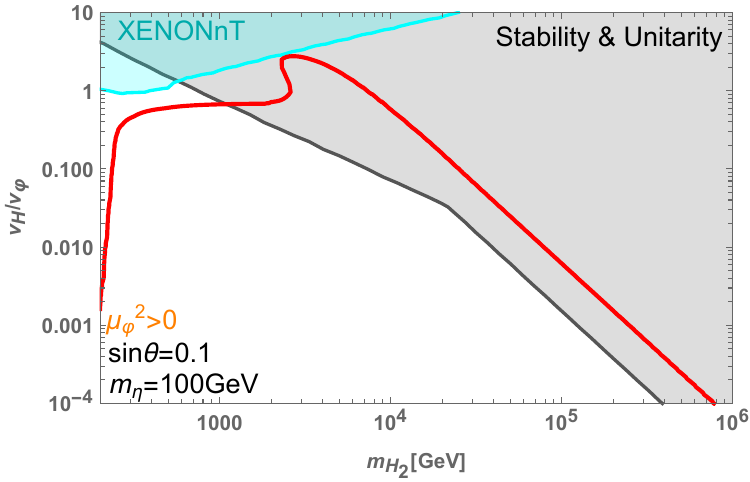}
\includegraphics[width=0.48\linewidth]{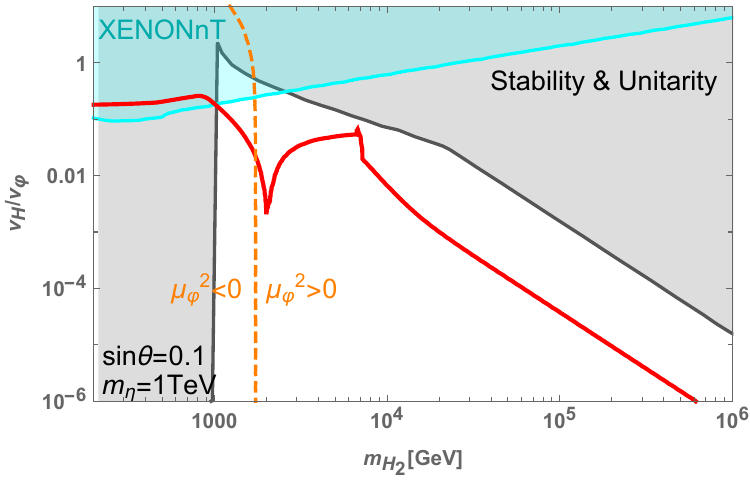}
\includegraphics[width=0.48\linewidth]{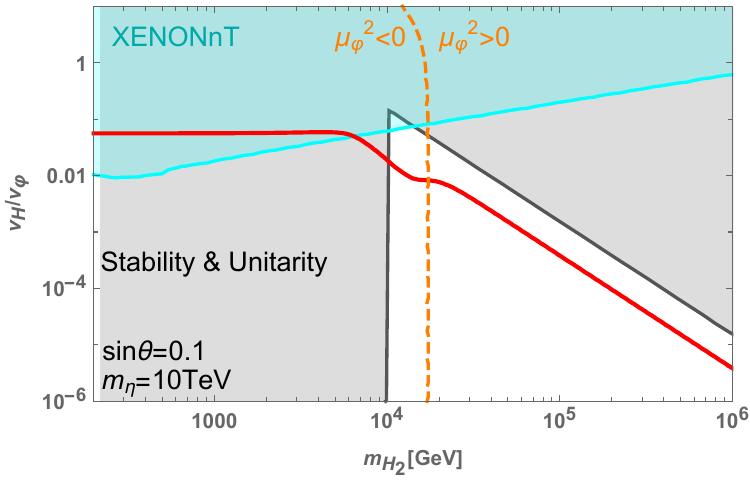}	
\includegraphics[width=0.48\linewidth]{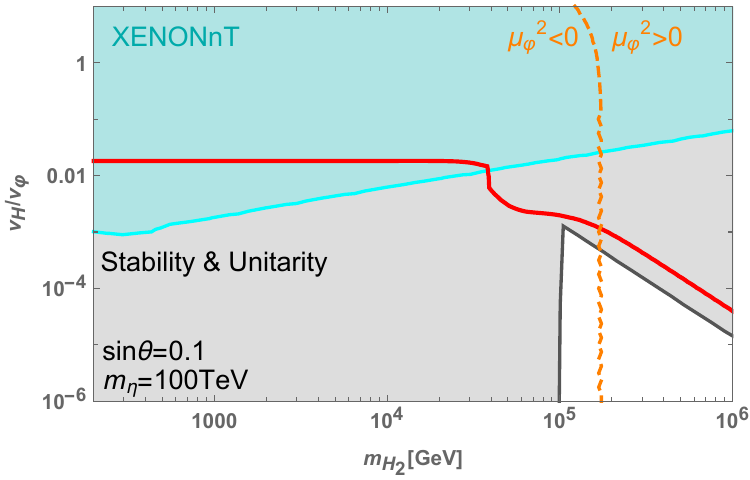}		
\hspace{0.5cm}
\caption{ Allowed region for $m_{H_2}$ vs $v_H/v_\varphi$. We take $\sin\theta=0.1$.  From top-left to bottom-right panels, we take $m_{\eta}=10^2, 10^3, 10^4,10^5$ GeV, respectively. The red line corresponds to the DM relic abundance which is consistent with the Planck data, $\Omega h^2=0.12 \pm 0.001$. The gray region is excluded by vacuum stability and perturbative unitarity conditions. 
The orange dashed line corresponds to $\mu_\varphi^2 = 0$. In the left (right) side of this orange line, one has
$\mu_\varphi^2 < 0$ $(\mu_\varphi^2 > 0)$, namely without (with) spontaneous breaking of $U(1)_X$. 
Cyan region is ruled out by DM direct detection constraint by XENONnT. }  \label{mH2VSvphi}
\end{figure}

In Fig.~\ref{mH2VSvphi}, we show the prefered paramter space in the $(m_{H_2}, v_H/v_\varphi)$ plane with 
the mixing angle $\sin\theta=0.1$ and $m_{\eta}=10^2,~10^3,~10^4,~10^5$ GeV, respectively.
The gray region is ruled out by the validation of the perturbative calculations and stability of the scalar potential.
The red solid line provides the correct DM relic density.
Cyan region is constrained by dark matter direct detection scattering process.
In the case of $m_\eta=1$ TeV and $10$ TeV and near the resonance  $m_{H_2}\sim m_\eta/2$, 
DM annihilations into a pair of $W, Z$ boson contribute significantly  to the DM relic density. 
Above the $H_2$ resonance, $\eta\eta \to H_1H_1$ annihilation channel is significant to get the correct relic density. 
When $m_\eta =100$TeV, all the parameter space which satisfy the relic abundance is excluded by the stability and perturbativity conditions.
The orange dashed line in Fig.~2 represents $\mu_\varphi^2 = 0$. The right side of the orange line corresponds to 
the region $\mu_\varphi^2 > 0$, where $U(1)_X$ is broken spontaneously. 

Now we describe indirect detection bound in pNGB DM models with linear symmetry breaking.
Here we consider four different final states $\eta \eta \to b\bar{b}, W^+ W^-, ZZ$ and $H_1H_1$.
\begin{figure}
\centering
\includegraphics[width=0.48\linewidth]{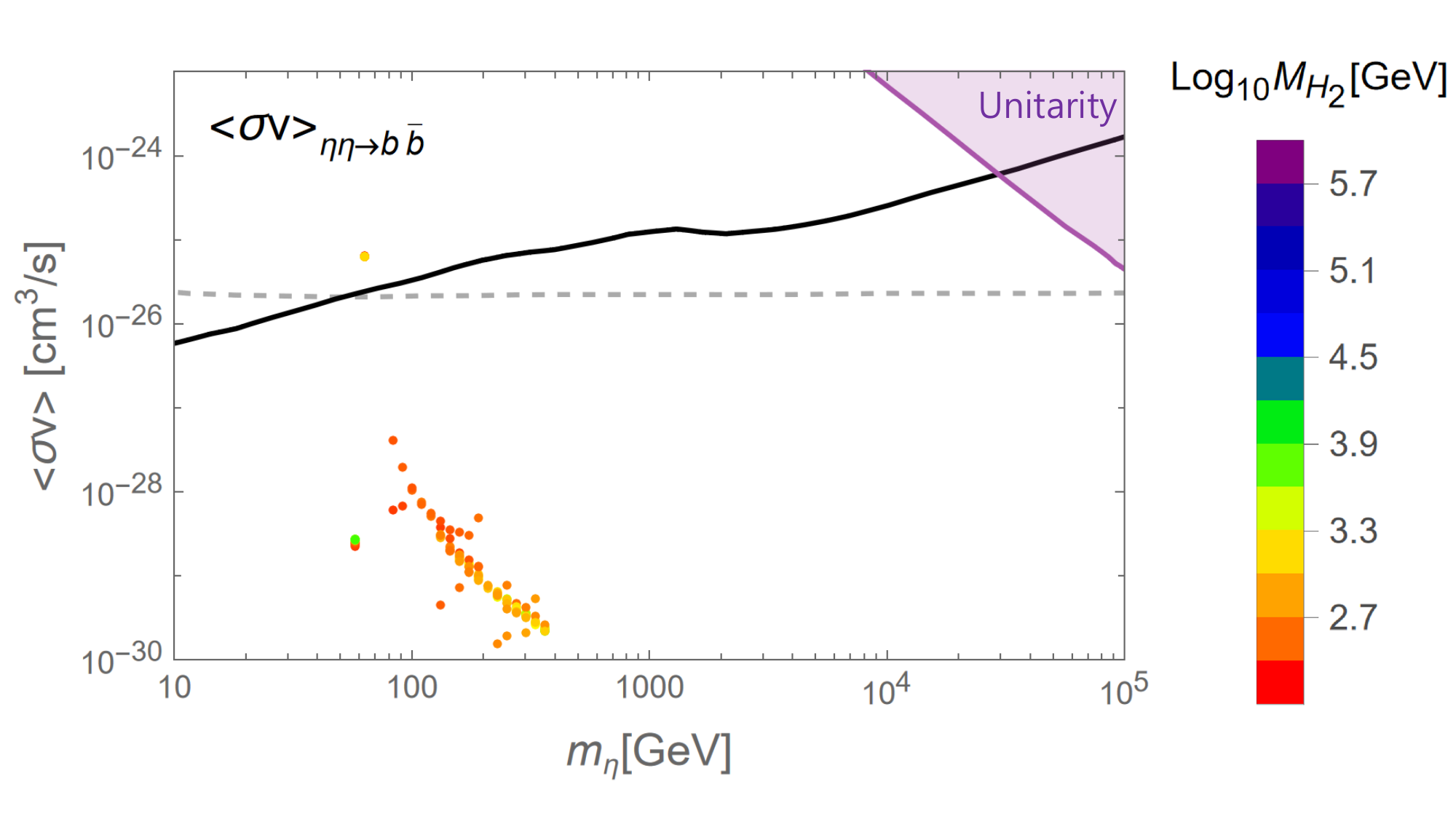}
\includegraphics[width=0.48\linewidth]{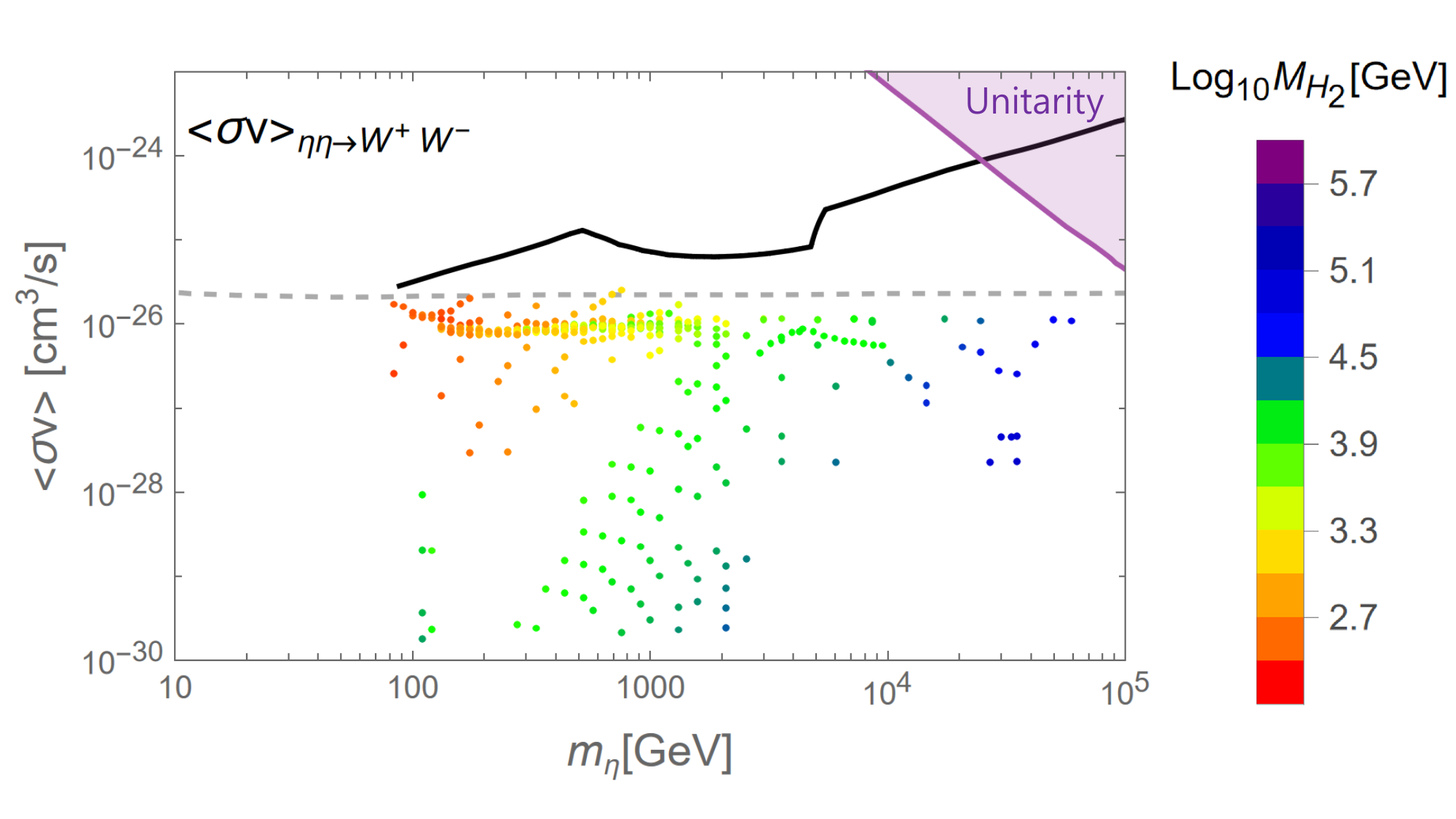}
\includegraphics[width=0.48\linewidth]{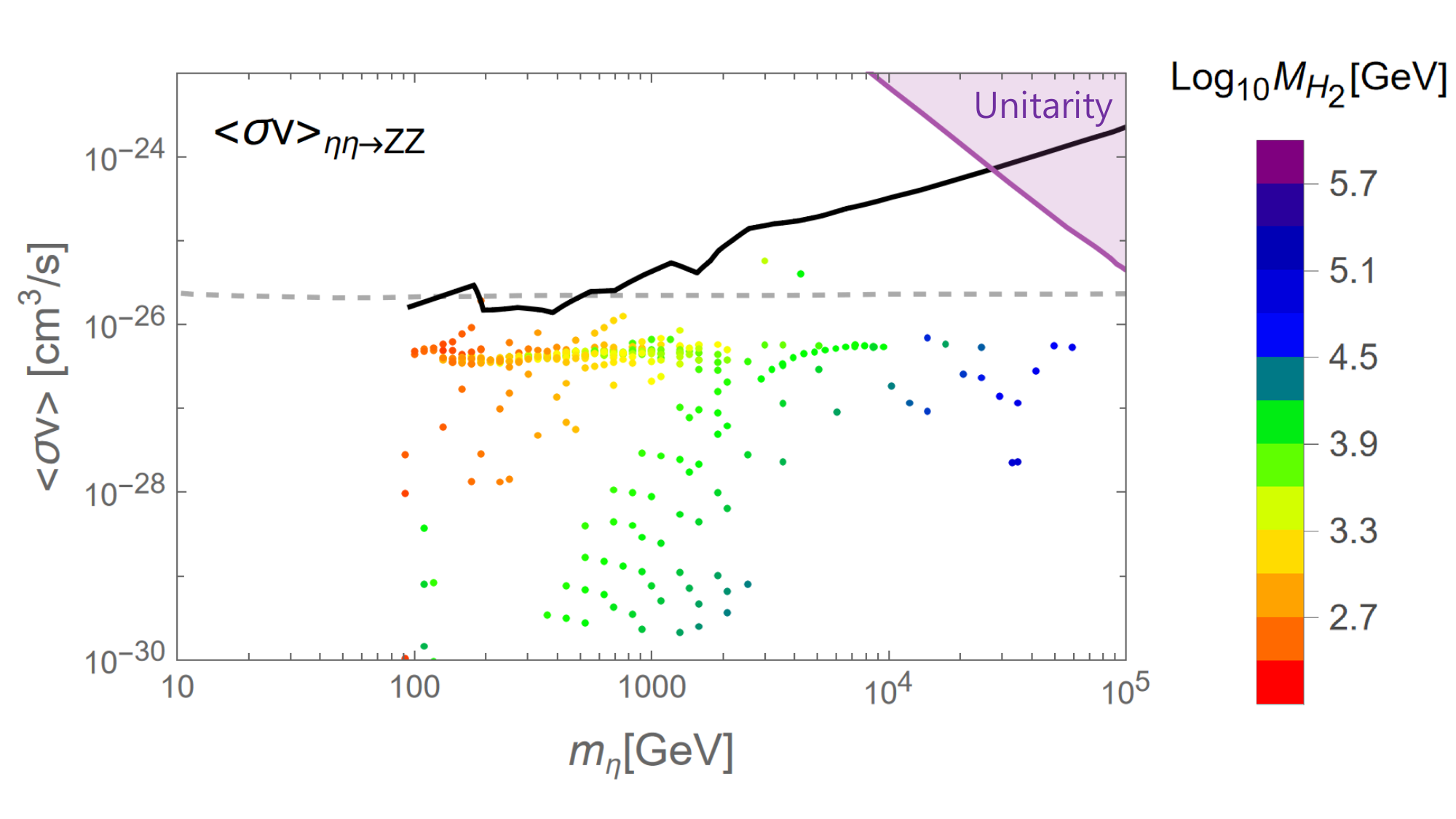}
\includegraphics[width=0.48\linewidth]{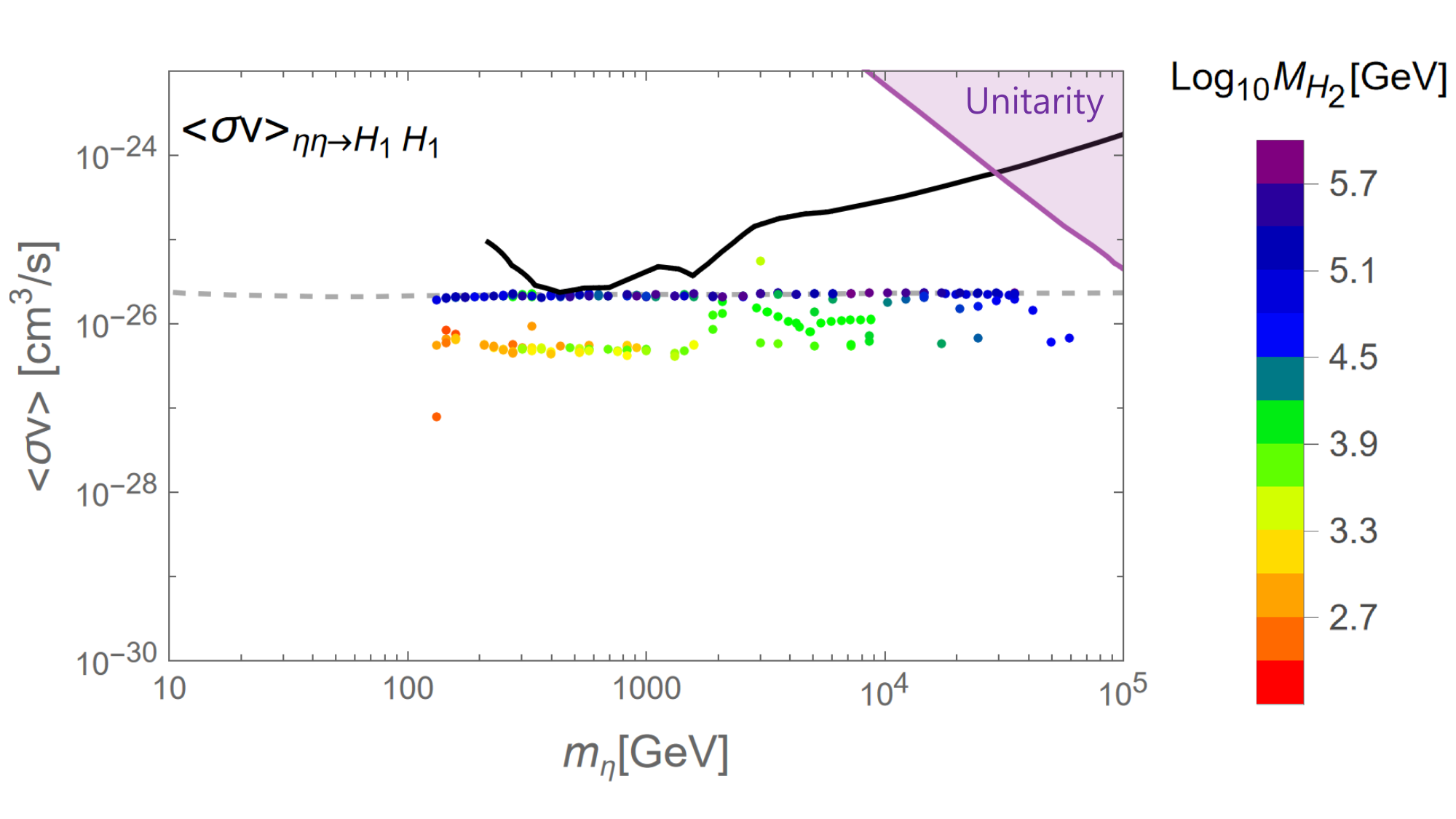}
\hspace{0.5cm}
\caption{Scatter plots in the  $m_{\eta}$ vs $\langle\sigma v\rangle$ plane. All the points are satisfied with DM relic density, stability condition in the scalar potential, perturbative constraints, Higgs invisible decay constraint and DM direct detection constraint depending on $M_{H_2}$. 
The value of the annihilation cross section that corresponds to the thermal production of DM is denoted with dashed gray line.Black lines are related to the current DM indirect detection bound from CMB, Fermi-LAT, H.E.S.S  and AMS-02. The shaded purple region is disfavored by the violation of unitarity in DM annihilation process.  }\label{indirect}
\end{figure}
In Fig.~\ref{indirect}, we show scatter plots for $m_{\eta}$ vs $\langle\sigma v\rangle$ plane.
Here we fix $\sin\theta=0.1$ since dependence of $\sin\theta$ in annihilation cross section is trivial.
Other free parameters are scanned over in the following ranges:
\begin{eqnarray}
0.115 \leq \Omega h^2 \leq 0.125, ~200{\rm GeV} \leq m_{H_2} \leq 10^6 {\rm GeV}, ~24.6 {\rm GeV} \leq v_\varphi \leq 2.5\times 10^8 {\rm GeV}.
\end{eqnarray}
Notice that all of the points in Fig.~\ref{indirect} satisfy the all constraints from stability of the potential, perturbative unitarity, Higgs invisible decay and DM direct detection searches.

In order to get the correct relic density, the required total annihilation cross section is around 
$3\times 10^{-26} {\rm cm^3/s}$. 
In most cases, the dominant DM pair annihilation channels are $W^+W^-, ZZ$ and $H_1H_1$.
When $m_{H_2}$ is very heavy $m_{H_2} \gg 1{\rm TeV}$, the DM relic density is mostly determined by 
DM annihilation into a pair of the SM Higgs boson.
The black line indicates the current DM scattering constraints from DM indirect detection searches.

As a result of the unitarity of the $S-$matrix, the 2-to-2 DM annihilation scattering cross section is bounded 
as follows \cite{Kamionkowski}: 
\begin{eqnarray}
\langle \sigma v_{rel} \rangle &<& \frac{4\pi (2j+1)}{m^2_{\rm DM} v_{rel} }  , 
\end{eqnarray}
where $j$ is the total angular momentum, and $v_{rel}\simeq 0.5$ corresponding to $x_f\approx 25$.
The purple region is excluded by the violation of the unitarity condition in DM annihilation cross section.

\section{Linear breaking with with quartic interaction of $\varphi$ ($\lambda_{\varphi}=0$)} \label{Linear:lambda0}

Let us consider the specific case of $\lambda_{\varphi}=0$ corresponding to the case of cosmological collider 
physics with Ref. \cite{Bodas:2020yho}.
This case should be treated separately from $\lambda_{\varphi} \neq 0$ discussed in the previous section,
since $\mu^2_\varphi$ should be negative in this case because of the stability of the scalar potential and 
the vanishing tadpole conditions are modified.  
In this case, there is no spontaneous symmetry breaking even if $v_\varphi \neq 0$, 
since there is no degeneracy in vacuum when we ignore the linear breaking term. 
Therefore DM $\eta$ is no longer pNGB in this case, and $\rho$ and $\eta$ masses will be degenerate 
when the Higgs portal coupling $\lambda_{H\varphi}$ is turned off.
  
The mass spectrum and couplings between DM and $H_i$ are changed from those in the previous section 
for $\lambda_{\varphi} > 0$.
The second order derivatives of the scalar potential are given by the following $3\times 3$  matrix:
\begin{align}
\partial_i \partial_j V &= 
\begin{pmatrix}
\lambda_H v_H^2 & \lambda_{H\varphi} v_H v_\varphi & 0 		\\
\lambda_{H\varphi} v_H v_\varphi  & m_\eta^2  & 0\\
0 & 0 & m_\eta^2 
\end{pmatrix} 	\label{deldelV:lamb0}
\end{align}
in the $(h,\rho,\eta)$ basis.  
We can see that masses of $H_2$ and $\eta$ fields becomes degenerate if the Higgs portal coupling  
$\lambda_{H\varphi}$ is turned off. And this degeneracy is lifted when $\lambda_{H\varphi}$ is turned on
\footnote{Strictly speaking, the Higgs portal coupling $\lambda_{H\varphi}$ was set to zero  in Ref. 
\cite{Bodas:2020yho}.  In such a case one has to consider gravitational production of DM which however 
goes  beyond the scope of this work. Therefore we shall assume that the Higgs coupling is nonzero, and assume 
$\lambda_\varphi = 0$ only in this Sections. }.

The modified dimensionful couplings $\{\lambda_{\eta\eta H_1},\lambda_{\eta\eta H_2}\}$ are
\begin{eqnarray}
\lambda_{\eta\eta H_1} &=&  -\frac{\left( m^2_{H_2} - m^2_{H_1}\right) \sin\theta\cos^2\theta}{v_\varphi},\\
\lambda_{\eta\eta H_2} &=& -\frac{\left( m^2_{H_2} - m^2_{H_1}\right) \sin^2\theta\cos\theta}{v_\varphi} .
\end{eqnarray}
Then, the tree-level direct detection scattering amplitude becomes
\begin{align}
\mathcal{M} &=  \frac{f_N m_N}{v_H v_\varphi }  \frac{ \left( m^2_{H_2} -m^2_{H_1}  \right) \left( m^2_{H_2} \sin\theta\cos^3\theta + m^2_{H_1}  \cos\theta\sin^3\theta   \right)    }{m^2_{H_1} m^2_{H_2}  }  \bar{u}_N u_N \nonumber\\
&\simeq \frac{f_N m_N}{v_H v_\varphi } \frac{ \left( m^2_{H_2} -m^2_{H_1}  \right)  \sin\theta    }{m^2_{H_1}  } \bar{u}_N u_N.
\end{align}
In the second line, we assume that $\sin\theta \ll 1$ and $\cos\theta \simeq 1$.
Note that there is no $t = q^2$ dependence in this case.
Then the spin-independent scattering process is
\begin{align}
\sigma_{\rm SI} &\simeq \frac{f^2_N}{4\pi }  \frac{m^4_N \sin^2\theta }{m^4_{H_1}  v^2_H v^2_\varphi}\frac{\left( m^2_{H_2} -m^2_{H_1} \right)^2}{\left(m_N+m_\eta\right)^2}.
\end{align}

\begin{figure}
\centering
\includegraphics[width=0.6\linewidth]{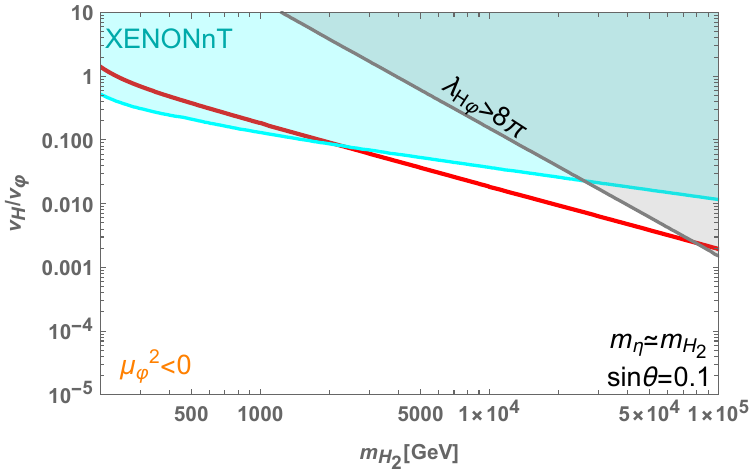}
\hspace{0.5cm}
\caption{ Allowed region for DM mass $m_{H_2}$ vs $v_H/v_\varphi$. We use $\sin\theta=0.1$. $m_{H_2}$ and $m_\eta$ are almost degenerate from Eq.\eqref{deldelV:lamb0}.
The red line corresponds to the DM relic abundance which is consistent with the Planck data, $\Omega h^2=0.12 \pm 0.001$ at $1\sigma$ C.L.. The gray area is excluded by perturbativity  of $\lambda_{H\varphi}$. All the parameter 
space satisfies the condition $\mu^2_\varphi  <0$ (the vacuum stability condition in case of $\lambda_\varphi = 0$). Lastly, the cyan region is ruled out by DM direct detection bound from XENONnT.  } \label{metaVSvphi:lambda0}
\end{figure}

In Fig.\ref{metaVSvphi:lambda0}, we note that the DM mass is allowed from a few TeV to $70$TeV, 
taking into account observed DM relic density.
The dominant annihilation cross section comes from $\eta\eta \to W^+ W^-, ZZ, H_1 H_1$.
When $m_\eta$ is below $2.7$TeV, the parameter space which satisfies with relic abundance is ruled out by 
XENONnT.   When $m_\eta$ is heavier than around $70$TeV, the region is excluded by the perturbative condition 
of $\lambda_{H\varphi}$.  All the parameter space in Fig.~\ref{metaVSvphi:lambda0} satisfies $\mu^2_\varphi  <0$.

All the dominant DM annihilation channels are in the s-wave.
For the allowed DM mass, the ratio for the dominant annihilation channels are almost the same, which is given by
\begin{align}
\langle \sigma v \rangle_{W^+W^-}:\langle \sigma v \rangle_{ZZ}:\langle \sigma v \rangle_{H_1,H_1} = 2:1:1.
\end{align}
Each DM annihilation cross section is below the thermal freeze-out annihilation, $3\times 10^{-26} {\rm cm^3/s}$.
This scenario cannot be testified through the current DM indirect detection bound. 

\section{Conclusions} \label{cons}
In this work, we studied pNGB DM model with a linear symmetry breaking term.
The model includes global dark $U(1)_X$ symmetry and a new SM-singlet complex scalar field charged 
under $U(1)_X$.  The imaginary part of new scalar field plays the role of pNGB dark matter after global $U(1)_X$ symmetry breaking,   both spontaneous and explicit breakings.  
Furthermore this model is interesting in the context of cosmological collider physics. 
Heavy particles with mass in the range $H \lesssim m_{\rm heavy} \lesssim 60 H$ can leave imprints 
on the non-Gaussianity at the level of  $f_{\rm NL} \sim O(0.01-10)$
 without Boltzmann supression, since heavy scalar current has the chemical potential form considered in 
Ref. \cite{Bodas:2020yho}. 

Detailed study of DM phenomenology of this model has been presented in this work.
We performed numerical analysis by combining constraints from the DM relic abundance, perturbative unitarity, 
vacuum stability, Higgs invisible decay, DM-nucleon scattering and DM indirect detection searches.
In contrast to the pNGB DM models with quadratic symmetry breaking, our model has a different cancellation 
mechanism in DM-nucleon scattering process, because DM mass in our model is generated from the linear breaking term.  In our case direct detection bounds becomes weaker in the limit $m_{H_2} \rightarrow m_{H_1}$, just like in the Higgs portal singlet fermion and vector DM cases.
We have examined the allowed parameter space even for $m_{H_2}=1$ PeV or heavier. 
Due to the perturbativity and the stability of the potential, DM mass cannot reach up to $50~{\rm TeV}$.
Also, DM mass cannot exceed heavy CP-even Higgs boson  due to the condition of 
$\lambda_\varphi \geq 0$.

For the special case of $\lambda_{\varphi}=0$, which was studied in Ref. \cite{Bodas:2020yho}  for 
cosmological collider physics, global $U(1)_X$ dark symmetry is no longer spontaneously broken, and 
heavy CP-even dark Higgs scalar boson and DM $\eta$  are nearly degenerate in their masses. 
Taking into account the observed DM relic density and bounds from XENONnT and perturbativity, the allowed 
DM mass lies between a few TeV and 70TeV.  

In the near future, some of parameter space in our model can be verified via DM indirect searches.
It would remain to be seen if such a heavy DM and CP-even scalar dark Higgs boson can leave any imprints 
on the cosmological collider data \cite{work_in_progress}.   They could leave  detectable imprints 
in the non-Gaussianity at the level of $f_{\rm NL} \sim O(0.01-10)$,  if the Hubble parameter during the inflation 
is as low as  $\sim O(1)$ TeV.

\acknowledgments
We are  grateful to Suro Kim and Sung Mook Lee for discussions on cosmological collider physics.
This work is supported by KIAS Individual Grants under Grants No. PG021403 (PK), and by National Research Foundation of Korea
(NRF) Research Grant NRF-2019R1A2C3005009 (JK, PK)


\bibliographystyle{JHEP}
\bibliography{refs.bib}

\end{document}